\documentclass[prb,twocolumn,showpacs]{revtex4}
\usepackage{amsmath}
\usepackage{graphicx}

\newcommand{\mysig}{\mathcal{S}}
\newcommand{\expe}{\mathrm{e}}

\begin{document}

\title{Vibrational sidebands and dissipative tunneling in molecular transistors}
\author{Stephan Braig$^1$ and Karsten Flensberg$^{1,2}$}
\affiliation{$^1$Laboratory of Atomic and Solid State Physics,
Cornell University, Ithaca, NY 14853\\ $^2$\O rsted Laboratory,
Niels Bohr Institute fAPG, Universitetsparken 5, 2100 Copenhagen,
Denmark.}
\date{\today}
\pacs{73.23.Hk, 73.63.-b, 85.65.+h}

\begin{abstract}
Transport through molecular devices with strong coupling to a
single vibrational mode is considered in the case where the
vibration is damped by coupling to the environment. We focus on
the weak tunneling limit, for which a rate equation approach is
valid. The role of the environment can be characterized by a
frictional damping term $\mysig(\omega)$ and corresponding
frequency shift. We consider a molecule that is attached to a
substrate, leading to frequency-dependent frictional damping of
the single oscillator mode of the molecule, and compare it to a
reference model with frequency-independent damping featuring a
constant quality factor $Q$. For large values of $Q$, the
transport is governed by tunneling between displaced oscillator
states giving rise to the well-known series of the Frank-Condon
steps, while at small $Q$, there is a crossover to the classical
regime with an energy gap given by the classical displacement
energy. Using realistic values for the elastic properties of the
substrate and the size of the molecule, we calculate $I$-$V$
curves and find qualitative agreement between our theory and
recent experiments on $C_{60}$ single-molecule devices.
\end{abstract}
\maketitle

\section{Introduction}

In the emerging field of single-molecule electronics there is a
large interest in transport through mesoscopic systems with strong
electron-phonon coupling. There has been a number of experiments
in which transport through a single molecule has been
reported.~\cite{reed97,park00,park02,lian02,smit02,zhit02} One
example is the series of experiments by Park {\it et
al.}~\cite{park00} where it was shown that the current through a
single $C_{60}$ molecule was strongly coupled to a single
vibrational mode. The single phonon mode was associated with the
motion of the molecule in the confining potential created by the
van-der-Waals interaction with the electrodes. Later, similar
devices with more complicated molecules were
investigated,~\cite{park02,zhit02} and they also showed excitation
spectra which may be associated with emission of vibrational
quanta.

Theoretically, there has been a large amount of work on the
problem of tunneling through a single level with coupling to
phonon modes. In many experimental realizations the tunnel
coupling to the leads is rather weak, and the transport is
dominated by the well-known Coulomb blockade effect. In this
regime, where the transport is sequential, the use of a rate
equation approach is appropriate rather than a coherent scattering
approach. Motivated by the above mentioned single-molecule
experiments, the rate equation approach has been used in a number
of recent papers.~\cite{boes01,mcca02,mitr03} Some of these
studies dealt with the issue of non-equilibrium phonon states, and
the possibility of having negative differential conductance in
such molecular systems.~\cite{boes01,mcca02} Physically, it is an
essential question how the excited vibrational levels are allowed
to relax, either through coupling to the environment, for example
the phonons or plasmons of the metal substrate, or by virtue of
the tunneling electrons.~\cite{aji03} In the case where the
relaxation of the vibrational mode is faster than the tunneling
rate one can assume an equilibrium phonon distribution.

\begin{figure}[pb]
\setlength{\unitlength}{1cm}
\begin{picture}
(6,4.5)(0,0) \put(-.5,0){\includegraphics[width=7cm]{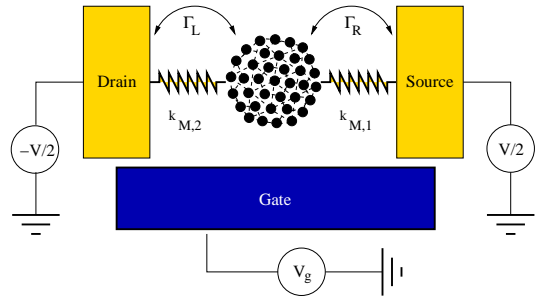}}
\end{picture}
\caption{Illustration of the system considered in this paper. The
molecule is attached to substrates, {\it e.g.}, by van-der-Waals
interactions, and the movement in this potential is modelled by
springs with spring constants $k_{M,1}$ and $k_{M,2}$. When an
electron hops onto the molecule, the force created by image
charges or local electric fields causes a shift of the equilibrium
position of the oscillator and consequently emissions of quanta.
The weights of the different final states are given by the
well-known Frank-Condon overlap factors. The main objective of
this paper is to consider the influence of damping of the
molecular motion by emission of phonons into the substrates. }
\label{fig:system}
\end{figure}

The coupling between the vibrational mode and the environment
depends strongly on which vibrational mode is considered. For
intra-molecular vibrations the lifetime can be very
long.~\cite{gure98,patt02} However, for the experiments of
Ref.~\onlinecite{park00}, it was suggested that the vibrational
motion was associated with a center of mass motion, which is
coupled to the environment more strongly as we discuss in this
paper. A sketch of the physical setup that we
consider is shown in Fig.~\ref{fig:system}.

In this paper, we assume that the tunneling rate is much smaller
than the rate of relaxation to other degrees of freedom. Hence,
the usual rate equation approach is applicable, and we can assume
that the phonons relax between each tunneling event according to a
thermal boson distribution. We study a model of one single
molecular orbital with strong Coulomb repulsion coupled to a
dissipative environment. The dissipation is caused by coupling to
phonon modes of the electrodes as well as electromagnetic modes
and it is represented by a bath of harmonic oscillators. The
description is thus similar to the well-known theory of Coulomb
blockade in an electromagnetic
environment.~\cite{naza89a,devo90,girv90,ingo91} However, there is
a difference in how the coupling to the environment appears. In
the electromagnetic environment case, the tunneling of an electron
results in a sudden displacement of the position of the charge,
while here the tunneling results in a sudden appearance of a force
on the oscillator. For this reason, we go through the derivations
in some detail and derive a general formula for the $I$-$V$
curves. This general result does not depend on the nature of the
environment but we then specialize to two cases. We consider a
molecule attached to a substrate, using a continuum model for the
substrate, and compare our result to a reference model featuring
frequency-independent damping and quality factor. The $I$-$V$
curves, as a function of the elastic parameters of the substrate
and the size of the molecule, feature quite different line shapes
as compared to the assumption of constant friction.

To get a simple estimate of the importance of the
coupling to the substrate, consider a model where the molecule
position $x$ is coupled to a one-dimensional substrate through a
spring with spring constant $k_M$. For small substrate
displacements, force balance gives that
\begin{equation}
-k_M x \approx v_s^2\rho_{1\mathrm{D}}\left(\frac{\partial u(z)}{\partial
z}\right)_{z=0},
\end{equation}
where $u(z)$ is the substrate displacement, $\rho_{1\mathrm{D}}$ is the 1D mass
density, and $v_s$ is the sound velocity. At a given frequency
 $\omega$, the outgoing soundwaves are $u(z,\omega)=a e^{i
z\omega/v_s}$, where $a$ is a constant. Finding $a$ from (1), we
can insert it into the equation of motion for $x$ to obtain the
quality factor $Q$ at the resonance frequency $\omega_0$:
\begin{equation}
Q=\frac{m_0 \rho_{1\mathrm{D}} v_s\omega_0^3}{k_M^2}
=\frac{\rho_{1\mathrm{D}} v_s}{m_0 \omega_0}
=\frac{m}{m_0},
\end{equation}
where $m_0$ is the molecule mass with $\omega_0^2=k_M/m_0$, and
$m=\rho_{1\mathrm{D}} v_s/\omega_0$ is the mass of a wavelength
long piece of the substrate. With realistic parameters for a
$C_{60}$ molecule on a gold substrate, as was used in the
experiments of  Ref.~\onlinecite{park00}, the quality factor is
between 1 and 10, and therefore we expect the broadening to be
substantial. This furthermore confirms the assumption that, for
this type of molecular device, relaxation through the environment
is much faster than through tunneling.

The paper is organized as follows. The model Hamiltonian is
defined in Section~\ref{sec:model}, and in Section~\ref{sec:rate}
we derive an expression for the current from rate equations. The
function that describes the tunneling density of states is then
solved in absence of the dissipative environment in
Section~\ref{sec:without} and with coupling to the environment in
Section~\ref{sec:with}. We discuss different models for the
dissipative coupling in Section~\ref{sec:sigmamodel}, where we
also discuss the physical implications. Section~\ref{sec:iv}
contains examples of $I$-$V$ curves, and finally, a summary as
well as a comparison with the experiments of
Ref.~\onlinecite{park00} can be found in Section~\ref{sec:sumdis}.

\section{Model Hamiltonian}

\label{sec:model}

We consider a model of one single spin-degenerate molecular level
coupled to two leads (generalization to more molecular levels is
straightforward). The single level is coupled to the vibrational
mode of the molecule through the charge on the dot. The coupling
between the oscillator and the environment is included as a linear
coupling to a bath of harmonic oscillators in the spirit of the
theory by Caldeira and Leggett.~\cite{cald83} The model
Hamiltonian then reads
\begin{equation}
H=H_{LR}^{{}}+H_{D}^{{}}+H_{B}^{{}}+H_{DB}^{{}
}+H_{\mathrm{bath}}^{{}}+H_{B\mathrm{bath}}^{{}}+H_{T}^{{}},
\label{Hstart}
\end{equation}
with
\begin{subequations}
\label{H}
\begin{align}
H_{LR}^{{}}  &
=\sum_{k\sigma,\,\alpha=L,R}\xi_{k\alpha}^{{}}c_{k\sigma,\alpha}^{\dagger
}c_{k\sigma,\alpha}^{{\vphantom{\dagger}}}\\
H_{D}^{{}}  &  =\sum_{\sigma}\xi_{0}^{{}}d_{\sigma}^{\dagger}d_{\sigma}^{{\vphantom{\dagger}}
}+Un_{d\uparrow}n_{d\downarrow},\\
H_{\mathrm{B}}^{{}}  &  =\frac{p_{0}^{2}}{2m_{0}^{{}}}+\frac{1}{2}m_{0}^{{}
}\omega_{0}^{2}x_{0}^{2},\\
H_{DB}^{{}}  &  =\lambda x_{0}^{{}}\sum_{\sigma}d_{\sigma}^{\dagger}d_{\sigma
}^{{}},\\
H_{\mathrm{bath}}^{{}}  &  =\sum_{j}\left(  \frac{p_{j}^{2}}{2m_{j}}+\frac
{1}{2}m_{j}\omega_{j}^{2}x_{j}^{2}\right)  ,\\
H_{B\mathrm{bath}}^{{}}  &
=\sum_{j}\beta_{j}^{{}}x_{j}^{{}}x_{0}^{{}}, \label{HBbath}
\end{align}
where the $c_{k\sigma,\alpha}^{\dagger },\,
c_{k\sigma,\alpha}^{{\vphantom{\dagger}}}$ and the
$d_{\sigma}^{\dagger},\,d_{\sigma}^{{\vphantom{\dagger}}}$ are
creation and annihilation operators for the leads and the dot,
respectively, $x_{0}^{{}}$ is the oscillator degree of freedom,
$\{x_{j}\}$ describes the set of environment degrees of freedom
and $m_{j}$, $\omega_{j}$ their respective masses and frequencies,
$\xi_{0}$ is the onsite energy, and $U$ is the Coulomb interaction
on the molecule. The coupling constants between
electron-oscillator and oscillator-bath are $\lambda$ and
$\{\beta_{j}\},$ respectively. The lead electron energies are
given by
\end{subequations}
\begin{equation}
\xi_{k\alpha}=\varepsilon_{k\alpha}-\mu_{\alpha},
\end{equation}
where $\mu_{\alpha}^{{}}$ is the chemical potential of lead
$\alpha$. Finally, the tunnel Hamiltonian is
\begin{equation}
H_{T}^{{}}=\sum_{k\sigma,\,\alpha=L,R}t_{k\alpha}^{{}}(c_{k\sigma,\alpha}^{\dagger
}d_{\sigma}^{{}}+d_{\sigma}^{\dagger}c_{k\sigma,\alpha}^{{}}).
\end{equation}
The tunneling amplitudes could in principle also depend on the
oscillator position. In the experimental realizations in
Refs.~\onlinecite{park00,park02,lian02}, this is probably a small
effect since the oscillator length is of order a few pm whereas
the tunneling matrix element changes on the scale of nm. For
simplicity, we do not take any such non-linear effects into
account, but we note that a position dependence of the tunneling
amplitudes, for example of the form $\exp(-x_0/\ell_{t})$, could
easily be included in the present formalism.

The force acting on the charged molecule, represented by the term
$H_{DB}$, is caused by electric fields originating from either static
impurity charges or image charges. Since this force on the molecule
is counteracted by a force on these charges and, hence, on the
environment, we should in principle also include the interaction
between the environment coordinates and the charge on the
molecule. This would in fact lead to a qualitatively different
behavior since Ohmic dissipation is cut off at frequencies
smaller than the inverse size of the total system, {\it i.e.}, the
inverse range of the interaction between the charge on the molecule
and the charges in the environment. This interesting subtlety was
pointed out in Ref.~\onlinecite{ard95}. However, since the
van-der-Waals interaction between molecule and the substrate is
short range compared to the electrostatic forces, we
will consider the force acting on the molecule as an external
quantity, which is thus not coupled to the dissipative
environment. We do note, however, that including such a coupling
would in fact lead to a small discontinuity at the beginning of
each step in the $I$-$V$ curve. The experimental data of, {\it
e.g.}, Ref.~\onlinecite{park00} does not seem to suggest such a
discontinuity, and we therefore specialize to the case where the
environment coordinates are unaffected by the charge on the
molecule.

We now want to relate the coupling constants in the boson-bath
coupling to the finite damping of the vibrational mode, which can
be accomplished by studying the classical equations of motion.
After removing the bath degrees of freedom, thereby neglecting the
term $H_{DB}^{{}}$ that will be removed by a unitray
transformation below, we end up with the following equation of
motion in the frequency domain:
\begin{eqnarray}
[\omega^{2}-\omega_0^{2}-\mysig(\omega)]x_{0}(\omega)=0,\label{xeqmfreq}
\end{eqnarray}
where we have defined
\begin{eqnarray}
\mysig(\omega)=\frac1{m_0}\sum_j\frac{\beta_j^2}{m_j}
\frac{1}{(\omega+i\eta)^2-\omega_j^2}, \label {mysigdef}
\end{eqnarray}
which is complex in general and gives rise to frictional damping
and a frequency shift of the bare frequency $\omega_0$.
In Section~\ref{sec:realistic}, we will explicitly calculate
$\mysig(\omega)$ for the case of a molecule attached to a
semi-infinite substrate.

We eliminate the coupling term $H_{DB}^{{}}$ of the
Hamiltonian (\ref{H}) by a unitary transformation similar to the
one used in the independent boson model,~\cite{mahan} at the cost
of introducing displacement operators in the tunneling term.
However, since we are dealing with a somewhat more complicated system due
to the coupling to the bosonic bath, the unitary transformation in
Ref.~\onlinecite{mahan} has to be generalized. We define the
transformation
\begin{equation}
\tilde{H}=SHS^{\dagger},\quad S=e^{-iAn_{d}^{{}}},\quad
A=p_{0}^{{}}\ell +\sum_{j}p_{j}^{{}}\ell_{j}^{{}}, \label{Sdef}
\end{equation}
where $n_{d}^{{}}=\sum_{\sigma}d_{\sigma}^{\dagger}d_{\sigma}$. Using that
\begin{equation}
\tilde{x}_{0}^{{}}=x_{0}^{{}}-\ell
n_{d}^{{}},\quad\tilde{x}_{j}^{{}}
=x_{j}^{{}}-\ell_{j}^{{}}n_{d}^{{}},
\end{equation}
it is a matter of simple algebra to show that the linear coupling
term $H_{DB}^{{}}$ cancels if we set
\begin{equation}
\ell=\frac{\lambda}{m_{0}^{{}}[\omega_{0}^{2}+\mysig(0)]},\quad\ell_{j}^{{}}=\frac
{-\ell\beta_{j}^{{}}}{m_{j}^{{}}\omega_{j}^{2}},\quad\label{Sdef2}
\end{equation}
and the Hamiltonian then transforms into
\begin{equation}
\tilde{H}=H_{LR}^{{}}+\tilde{H}_{D}^{{}}+H_{B}^{{}}+H_{\mathrm{bath}}^{{}
}+H_{B\mathrm{bath}}^{{} }+\tilde{H}_{T}^{{}}, \label{Huni}
\end{equation}
where
\begin{equation}
\tilde{H}_{T}=\sum_{k\sigma,\,\alpha=L,R}t_{k\sigma,\alpha}^{{}}\left(
c_{k\sigma,\alpha}^{\dagger}e^{iA}d_\sigma^{{}}+d_{{\sigma}}^{\dagger}
e^{-iA}c_{k\sigma,\alpha}^{{}}\right) \label{tildeHT}
\end{equation}
and
\begin{equation}
\tilde{H}_{D}^{{}}=\varepsilon_{0}^{{}}\sum_\sigma
d_{{\sigma}}^{\dagger}d_\sigma^{{}}
+\tilde{U}n_{d\downarrow}^{{}}n_{d\uparrow}^{{}},
\quad\varepsilon_{0}^{{}}=\xi_{0}^{{}}-\frac{1}{2}\lambda\ell.
\end{equation}
Here, $\tilde{U}=U-\lambda\ell$ is the Coulomb repulsion modified
by the phonon mediated interaction. For a weak Coulomb
interaction, this can result in a negative effective $U$, which
was discussed in Ref.~\onlinecite{alex02}.

\section{Rate equations and current formula}

\label{sec:rate} We derive an expression for the current in
the weak tunneling limit using the usual kinetic equation
approach. As mentioned in the Introduction, the most important
assumption here is that the tunneling rate is much smaller than
all other time scales, which means that we can assume the
vibrational degrees of freedom and the Fermi seas in the two
electrodes to be in equilibrium at all times. For simplicity, we
consider only two charge states and therefore let $U=\infty$,
which leaves us with only three states: empty, and occupied by
either spin up or down. The probabilities for the three states are
denoted $P_{0} $, $P_{\uparrow}$, and $P_{\downarrow}$,
respectively. The rate equations are
\begin{equation}
\left(
\begin{array}
[c]{ccc}
-2\Gamma_{10}^{{}} & \Gamma_{01}^{{}} & \Gamma_{01}^{{}}\\
\Gamma_{10}^{{}} & -\Gamma_{01}^{{}} & 0\\
\Gamma_{10}^{{}} & 0 & -\Gamma_{01}^{{}}
\end{array}
\right)  \left(
\begin{array}
[c]{c}
{P}_{0}^{{}}\\
{P}_{\uparrow}^{{}}\\
{P}_{\downarrow}^{{}}
\end{array}
\right)  =0,
\end{equation}
which combined with the condition
$P_{0}^{{}}+P_{\downarrow}^{{}}+P_{\downarrow }^{{}}=1$ has the
solution
\begin{equation}
P_{0}^{{}}=\frac{\Gamma_{01}^{{}}}{\Gamma_{01}^{{}}+2\Gamma_{10}^{{}}},\quad
P_{\downarrow}^{{}}=P_{\uparrow}^{{}}=\frac{\Gamma_{10}^{{}}}{\Gamma_{01}^{{}
}+2\Gamma_{10}^{{}}},
\end{equation}
where $\Gamma_{10}$ is the tunneling rate for tunneling from the
empty state to a singly occupied state, and $\Gamma_{01}$ is the
rate for the reverse process. Since the electron can tunnel out of
both left and right leads, both rates have left and right
contributions: $\Gamma_{ij}=\Gamma_{ij}^{L}+\Gamma_{ij}^{R}.$ The
tunneling rates are calculated using Fermi's Golden Rule, thereby
treating $\tilde{H}_T$ of Eq.~(\ref{tildeHT}) as the perturbation
and assuming a thermal equilibrium distribution of the lead
electrons and the phonon bath. Following standard derivations, we
obtain
\begin{subequations}
\label{Gamma}
\begin{align}
\Gamma_{10}^{\alpha}  &
=\Gamma_{\alpha}^{{}}\int\frac{d\omega}{2\pi}
F(\omega)n_{\alpha}^{{}}(\varepsilon_{0}+\omega),\label{Gamma10}\\
\Gamma_{01}^{\alpha}  &
=\Gamma_{\alpha}^{{}}\int\frac{d\omega}{2\pi
}F(-\omega)(1-n_{\alpha}^{{}}(\varepsilon_{0}+\omega)),
\label{Gamma01}
\end{align}
where we have defined the function
\end{subequations}
\begin{align}
F(\omega)  &  =\int_{-\infty}^{\infty}dt\,e^{i\omega t}F(t),\,\,\,\,\, F(t)
  =\langle e^{iA(t)}e^{-iA}\rangle, \label{Ftdef}
\end{align}
in addition to the Fermi distributions of the two leads,
$n_{\alpha}(\varepsilon)$=$(e^{\beta(\varepsilon-eV_{\alpha})}+1)^{-1}$,
and the bare rates
$\Gamma_{\alpha}^{{}}$=$2\pi\sum_{k}|t_{k\alpha}|^{2}\delta(\xi_{k})$.
The function $F$ has the properties
\begin{equation}
F(\omega)=F(-\omega)e^{\beta\omega},\quad\int_{-\infty}^{\infty}\frac{d\omega
}{2\pi}F(\omega)=1. \label{Fwmw}
\end{equation}
We can use Eq.~(\ref{Fwmw}) to show that the rates in
Eq.~(\ref{Gamma}) can be written as
\begin{equation}
\label{Gamman}
\Gamma_{10}^{\alpha}=\Gamma_{\alpha}^{{}}\tilde{n}_{\alpha}^{{}},\quad
\Gamma_{01}^{\alpha}=\Gamma_{10}^{\alpha}e^{\beta(\varepsilon_{0}-eV_{\alpha
})},
\end{equation}
where we defined
\begin{equation}
\tilde{n}_{\alpha}^{{}}=\int\frac{d\omega}{2\pi}F(\omega)n_{\alpha}^{{}}
(\omega+\varepsilon_{0}).
\end{equation}

The current through the molecule is now given by
\begin{equation}
I=-e(2P_{0}^{{}}\Gamma_{10}^{L}-\left(
P_{\uparrow}^{{}}+P_{\downarrow}^{{} }\right)
\Gamma_{01}^{L})=2e\frac{\Gamma_{10}^{R}\Gamma_{01}^{L}-\Gamma
_{01}^{R}\Gamma_{10}^{L}}{\Gamma_{01}^{{}}+2\Gamma_{10}^{{}}}.
\end{equation}
Using Eq.~(\ref{Gamman}), this can also be written as
\begin{equation}
I=\frac{2e\Gamma_{L}^{{}}\Gamma_{R}^{{}}\tilde{n}_{R}^{{}}\tilde{n}_{L}^{{}
}\left(
e^{\beta(\varepsilon_{0}^{{}}-eV_{L}^{{}})}-e^{\beta(\varepsilon
_{0}^{{}}-eV_{R}^{{}})}\right)
}{\Gamma_{L}^{{}}\tilde{n}_{L}^{{}}(2+e^{\beta
(\varepsilon_{0}-eV_{L})})+\Gamma_{R}^{{}}\tilde{n}_{R}^{{}}(2+e^{\beta
(\varepsilon_{0}-eV_{R})})}. \label{I}
\end{equation}

\section{Without coupling to the environment}

\label{sec:without} We start by discussing the limit when the
oscillator is not coupled to the environment, which means that
thermal smearing dominates over dissipative broadening. However,
we still assume that the coupling is stronger than the tunnel
coupling so that the molecule equilibrates between each tunneling
event. This section is thus equivalent to the results in other
rate equation calculations, but for completeness and later
comparison we write down this limiting case.

The phonon average is performed assuming thermal equilibrium, and
we have
\begin{align}
F_{0}^{{}}(t) &  =\left\langle e^{ip_0(t)\ell}e^{-ip_0(0)\ell}\right\rangle
,\label{Ft}\\
&  =\exp\left\{  g\left(  e^{-i\omega_{0}t}-1\right)  (1+N)+g\left(
e^{i\omega_{0}t}-1\right)  N\right\}  ,\nonumber
\end{align}
where
\begin{equation}
g=\frac{1}{2}\left(  \frac{\ell}{\ell_{0}}\right)
^{2},\quad\ell_{0} ^{2}=\frac{1}{m_{0}^{{}}\omega_{0}},\quad
N=n_{B}^{{}}(\omega_{0}^{{} }),\label{gdef}
\end{equation}
\begin{figure}[ptb]
\setlength{\unitlength}{1cm}
\begin{picture}
(8,6.3)  \put(0,0){\includegraphics[width=8cm]{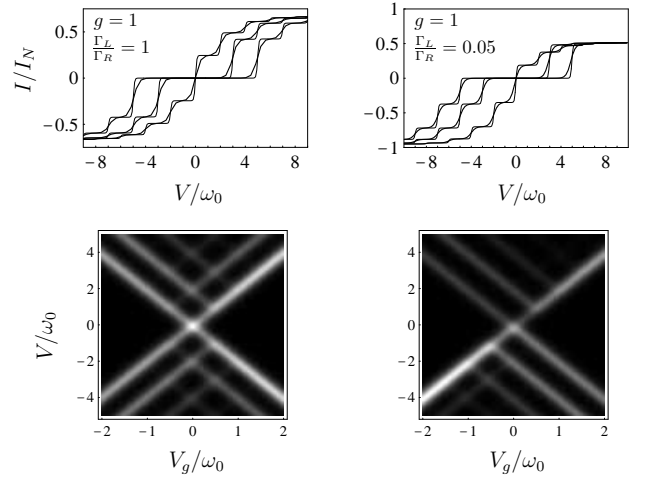}}
\end{picture}
\caption{Upper panels: current-voltage characteristics for a
device without coupling to the environment for symmetric
($\Gamma_{R}^{{}}=\Gamma_{L}^{{}}$) and asymmetric
($\Gamma_{R}^{{}}=0.05\Gamma_{L}^{{}}$) tunneling contacts. Lower panels: contour
plot of the differential conductance in the voltage-gate voltage
plane, where $eV_g=\varepsilon_0$. The curves have been calculated
using the analytic result Eq.~(\ref{I}), valid in the limit where
the lifetime broadening of the oscillator is negligible. The
temperature is $kT=0.1\omega_{0}^{{}}$ for the thick lines and in
the contour plots, while the thin lines are for
$kT=0.025\omega_{0}^{{}}$. The bias is applied symmetrically
$V_{L}=V/2=-V_{R}$ and in the $I$-$V$ curves we take $\varepsilon_0$
to be 0, 1.5, and 2.5 times $\omega_{0}^{{}}$. The current is
measured in units of
$I_N=e\Gamma_L^{{}}\Gamma_R^{{}}(\Gamma_L^{{}}+\Gamma_R^{{}})$.}
\label{fig:contour}
\end{figure}
Here, $g$ is an important parameter determined by the ratio of the
classical displacement length and the quantum mechanical
oscillator length. The evaluation of $F_{0}(\omega)$ from
Eq.~(\ref{Ft}) is equivalent to the independent boson
model,~\cite{mahan} and using the result from there we get
\begin{equation}
F_{0}^{{}}(\omega)=2\pi\sum_{n=-\infty}^{\infty}P_{n}(g)\delta(\omega
-n\omega_{0}),\label{DkT}
\end{equation}
where
\begin{equation}
P_{n}^{{}}(g)=\exp(-g\coth(b))e^{nb}I_{n}\left(  \frac{g}{\sinh(b)}\right)
,\quad b=\frac{\beta\omega_{0}^{{}}}{2},
\end{equation}
and $I_{n}^{{}}$ is the modified Bessel function of the first
kind. The finite-temperature result involves both positive and
negative values of $n$, corresponding to emission and absorbtion
of phonons, respectively. At zero temperature, this reduces to
having only positive values of $n$ because of the factor
$e^{n\beta\omega_{0}^{{}}/2}$, and hence only emission processes
are possible. In the limit $T\rightarrow0$, we thus have a series
of emission peaks at $\omega=n\omega_{0}$ for positive $n$ and
with weights given by the Poisson distribution $P_{n}\rightarrow
e^{-g}g^{n}/n!.$

The current can now be found from Eq.~(\ref{I}). In
Fig.~\ref{fig:contour}, we show examples of current-voltage
characteristics using Eq.~(\ref{I}) for symmetric and asymmetric
junctions. In the following, we study how the physics gets modified
by the coupling to the environment.

\section{With coupling to the environment}

\label{sec:with}

In presence of coupling to the environment, the evaluation of the
function $F(t)$ in Eq.~(\ref{Ftdef}) is in principle
straightforward since the Hamiltonian is quadratic in the oscillator and
bath degrees of freedom. We obtain
\begin{equation}
F(t)=\exp\left(  B(t)-B(0)\right)  ,\quad B(t)=\langle
A(t)A(0)\rangle_{0}, \label{Bt}
\end{equation}
where the operator $A$ is defined in
Eqs.~(\ref{Sdef}) and (\ref{Sdef2}). The expectation value
$\langle\dots\rangle_{0}^{{}}$ is to be evaluated with respect to
$\tilde{H}$ without the tunneling term. At this point, it is
convenient to use the fluctuation-dissipation theorem,
\begin{equation}
B(\omega)=-2\operatorname{Im}[B^{R}(\omega)](1+n_{B}^{{}}(\omega)),
\label{flucdis}
\end{equation}
to express $B(t)$ in terms of the corresponding retarded Green's function
\begin{equation}
B^{R}(t)=-i\theta(t)\langle\lbrack A(t),A(0)]\rangle_{0}.
\end{equation}
Here, $n_{B}(\omega)=(e^{\beta\omega}-1)^{-1}$ is the usual Bose
function. In order to find this retarded correlation function, we
define the following auxiliary Green's functions:
\begin{equation}
G_{\mathcal{O}}^{R}(t)  =-i\theta(t)\frac{1}{\ell}
\langle\lbrack \mathcal{O}(t),A(0)]\rangle,
\end{equation}
from which we obatin $B^{R}$ as
\begin{align}
B^{R}  &  =(\ell G_{p_0}^{R}+\sum_{j}\ell_{j}G_{p_j}^{R})\ell\nonumber\\
&
=\ell^{2}(G_{p_0}^{R}-\sum_{j}\frac{\beta_{j}^{{}}}
{m_{j}^{{}}\omega_{j}^{2}}G_{p_j}^{R}).\label{BR}
\end{align}
The equations of motion for these functions are in frequency
domain given by
\begin{equation}
\left(
\begin{array}
[c]{cc}%
\omega & -i/m_{0}\\
im_0\omega_{0}^{2} & \omega
\end{array}
\right)  \left(
\begin{array}
[c]{c}
G_{x_0}^{R}\\
G_{p_0}^{R}
\end{array}
\right)  =\left(
\begin{array}
[c]{c}
i\\
0
\end{array}
\right)  -\sum_{j}\left(
\begin{array}
[c]{c}
0\\
i\beta_{j}^{{}}G_{x_j}^{R}
\end{array}
\right),\label{Gxp}
\end{equation}
\begin{equation}
\left(
\begin{array}
[c]{cc}%
\omega+i\eta & -i/m_{j}^{{}}\\
im_j\omega_{j}^{2} & \omega+i\eta
\end{array}
\right)  \left(
\begin{array}
[c]{c}%
G_{x_j}^{R}\\
G_{p_j}^{R}%
\end{array}
\right)  =\left(
\begin{array}
[c]{c}%
\frac{-i\beta_{j}}{m_j\omega_{j}^{2}}\\
0
\end{array}
\right)  -\left(
\begin{array}
[c]{c}
0\\
i\beta_{j}^{{}}G_{x_0}^{R}
\end{array}
\right).\label{Gxpj}
\end{equation}
Solving this linear set of equations for the Green's functions and
inserting the results into Eq. (\ref{BR}), we obtain
\begin{equation}
B^{R}(\omega)=\frac{2g\bar{\omega}_{0}}{\omega^{2}-\bar{\omega}_{0}^{2}
-\bar{\mysig}(\omega)}\left(  1-\frac{\bar{\mysig}(\omega)}{\omega^2}\right),
\end{equation}
where we have defined $\bar{\mysig}(\omega)\equiv
\mysig(\omega)-\mysig(0)$ and the experimentally observable
renormalized frequency $\bar{\omega}_0^2\equiv
\omega_0^2+\mysig(0)$. Using (\ref{flucdis}), the function
$B(\omega)$ thus follows as
\begin{equation}
B(\omega)=-4g\;\frac{1+n_{B}(\omega)}{\omega^2}
\;\mathrm{Im}\left[\frac{\bar{\omega}_{0}^{3}}{\omega^{2}
-\bar{\omega}_{0}^{2}-\bar{\mysig}(\omega)]}\right],
\label{Bw}
\end{equation}
where now $g=\ell^2/2\ell_0^2$ is defined with respect to the renormalized
frequency $\bar{\omega}_0$, {\it i.e.} $\ell_0=1/m_0\bar{\omega}_0$.
This result can then be used to find
\begin{equation}
F(t)=\exp\left(\int_{-\infty}^{\infty}
\frac{d\omega}{2\pi} (\expe^{-i\omega t}-1)B(\omega)\right)
. \label{F1}
\end{equation}

Eq.~(\ref{F1}) is equivalent to the result for the Coulomb
blockade of a single tunnel junction with coupling to the
electromagnetic environment.~\cite{girv90,devo90} In the
Coulomb blockade problem, the tunneling density of states was
related to the impedance as seen from the junction, here the
$I$-$V$ characteristic is in a similar way related to the
frictional damping of the oscillator mode. In both cases, the low
energy form of the spectrum is a power law at low temperatures. At
small frequencies $\omega\ll\bar{\omega}_{0}^{{}}$ and zero
temperature, we get the following power law behavior:
\begin{equation}
F(\omega)\propto\omega^{\alpha-1},\quad\alpha=\frac{2g}
{\bar{\omega}_0\pi}\lim_{\omega\rightarrow
0}\left(\frac{\mathrm{Im}\bar{\mysig}(\omega)}{\omega}\right).
\label{Fpower}
\end{equation}
Furthermore, we use the trick by Minnhagen\cite{minn76} to find
the $F$-function as the solution of the integral equation
\begin{equation}
F(\omega)=\frac{1}{\omega}\int_{0}^{\omega}\frac{d\zeta}{2\pi} F(\zeta )
B(\zeta-\omega)(\zeta-\omega), \label{F1int}
\end{equation}
which is useful for the numerical evaluation of $F$.

\section{Models for $\mysig(\omega)$}
\label{sec:sigmamodel}

\subsection{Frequency-independent quality factor $Q$}
As a first attempt, we can assume $\mysig$ to be of the form
\begin{eqnarray}
\mysig(\omega)=\mysig(0)+i\frac{\omega\bar{\omega}_0}{Q},
\end{eqnarray}
which leads to a frequency-independent quality factor $Q$ of the
single vibrational mode, similar to the Ohmic dissipation model of
Caldeira and Leggett in Ref.~\onlinecite{cald83}. The model is
also similar to the Coulomb blockade problem of an LCR
circuit,~\cite{ingo91} which is described by the same formula. The
limit $Q\rightarrow\infty$ is seen to coincide with the results in
Section~\ref{sec:without} since in this limit
$B(\omega)\rightarrow
2\pi(1+n_{B}^{{}}(\omega))(\pm)\delta(\omega\pm\omega_{0})$, and
when this is inserted into Eq.~(\ref{F1}), we get (\ref{Ft}). We
also see that, for a critical value of $Q_{c}=2g/\pi$, the
function $F$ and, hence, the differential conductance change from
having a divergence at small energies to vanish at small energies.

\begin{figure}[ptb]
\setlength{\unitlength}{1cm}
\begin{picture}
(6,9)(0,0) \put(-1,0){\includegraphics[width=8cm]{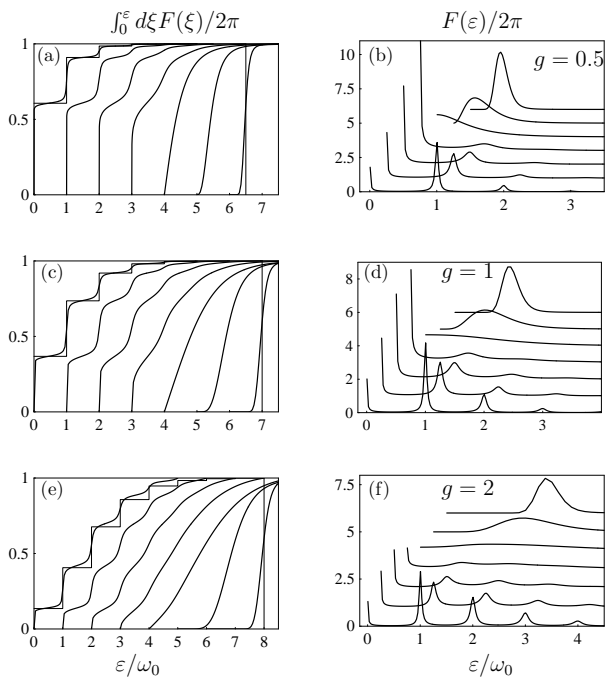}}
\end{picture}
\caption{ The function $F(\xi)$ and its integral for different
values of $g$ and frequency-independent $Q$. The curves have been
calculated from Eq.~(\ref{F1int}) at zero temperature. We take
$Q=20,10,5,2.5,2g/\pi ,0.1,0.01$, where $g=0.5$ in (a) and (b),
$g=1$ in (c) and (d), and $g=2$ in (e) and (f). The curves have
been displaced for clarity by multiples of (1,0) in (a), (c), and
(e), and by multiples of (0.25,1) in (b), (d), and (f) (largest
$Q$ to the left). For large $Q$, the integrated function goes to
the dissipationless results, where the step heights are given by
the Poisson distribution (leftmost staircase in (a), (c), and
(e)), whereas for small $Q$ it goes to a step function at
$\varepsilon=g\omega_0^{{}}$ (vertical line). Note also that the
differential conductance at the first step remains sharp while the
higher order steps are smeared when $Q>Q_{c}$.} \label{fig:Qg052}
\end{figure}
In Fig.~\ref{fig:Qg052}, we plot the function $F$ and its integral
for different values of $g$ and $Q$. It is clearly seen how the
increasing dissipation smears the Frank-Condon steps. For strongly
underdamped coupling to the environment, the steps are only weakly
smeared, and still visible even for $Q=2.5$. For the special value
of
\begin{equation}
Q=Q_{c}=\frac{2g}{\pi},
\end{equation}
the first step disappears and eventually for very small $Q$ the
function $F$ goes towards a delta function,
$F\rightarrow2\pi\delta(\omega-g\omega_{0})$. Physically, this
means that in the small $Q$ limit, the system relaxes immediately
to the classical state and tunneling is only possible by paying
the total classical energy cost of the displacement. To see this,
we rewrite $g\omega_{0}^{{}}$ in terms of the coupling constant
$\lambda$ and get
$\lambda\ell/2=\lambda^{2}/2m_{0}\omega_{0}^{2}$, which is the
classical energy for displacing the oscillator by increasing the
occupation $n_{d}$ by one.

The crossover to the classical regime occurs when the lifetime
of the oscillator, $\omega_{0}^{{}}/Q$, is comparable to the
Heisenberg uncertainty time associated with the classical
 energy of the displaced oscillator, {\it i.e.} when
(reinserting $\hbar$)
\begin{equation}
\frac{Q_q^{{}}}{\omega_0^{{}}}\equiv\frac{\hbar}{\lambda\ell/2}\Rightarrow
Q_{q}^{{}}=\frac{1}{g}\,.
\end{equation}
\begin{figure}[t]
\setlength{\unitlength}{1cm}
\begin{picture}
(6,5.15)(0,0) \put(-.5,0){\includegraphics[width=7cm]{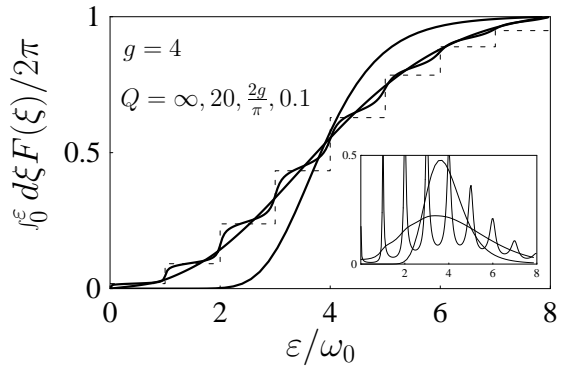}}
\end{picture}
\caption{ The integral of the function $F$ for $g=4$ and different
$Q$ at zero temperature. We take $Q=\infty$ (dashed line), 20,
$8/\pi$, 0.1. At $Q\lesssim Q_{c}^{{}}$ the steps disappear, while
for $Q\lesssim Q_{q}^{{}}$ the curves approach the classical
limit, which is a step function at $\varepsilon=4\omega_{0}^{{}}$.
The inset shows the $F$-function itself for the same parameters.}
\label{fig:Qg4}
\end{figure}
The disappearance of the steps, which happens at $Q_{c}^{{}}$, is
therefore different from the crossover to the classical regime.
This is shown in Fig.~\ref{fig:Qg4} where we plot the integral of
$F$ for $g=4$ for different values of $Q$. For $Q=20$, the steps
are only slightly broadened, while for $Q=Q_{c}^{{}}$, the steps are
almost fully broadened but the line still follows the quantum
behavior. Only for smaller $Q$ do we approach the classical
result, which is a step function at $g\omega_{0}^{{}}$.

\subsection{Coupling to a substrate}
\label{sec:realistic} We consider a molecule of mass $m_0$
attached to a substrate that extends over the semi-infinite
half-space $z\geq0$. The case of two substrates, as is shown in
Fig.~\ref{fig:system}, is a straightforward generalization and
will be discussed  at the end of the calculation. The 3D
Lagrangian density for the substrate is given by:~\cite{kn:ezawa}
\begin{eqnarray}
\mathcal{L}(\vec{r},t)&=&\frac12 \rho \left[\left(\partial_t \vec{u}\right)^2
-(v_l^2-2v_t^2)\left(\vec\nabla\vec{u}\right)^2\right.\nonumber\\
&&\qquad\left.-v_t^2(\vec\nabla\times\vec{u})^2-2v_t^2\frac{\partial u_i}{\partial x_j}
\frac{\partial u_j}{\partial x_i}\right],
\end{eqnarray}
where $v_l$ and $v_t$ are the longitudinal and transverse sound
velocities, and $\rho$ is the mass density.  This Lagrangian leads
to the following equation of motion:
\begin{eqnarray}
\partial_t^2\vec{u}
-v_l^2\vec\nabla(\vec\nabla \vec{u})+v_t^2\vec\nabla\times\vec\nabla\times\vec{u}=0
\end{eqnarray}
Having in mind a small molecule attached to the origin, the
assumption of cylindrical symmetry around the $z$-axis seems
reasonable. We define $u_r$ and $u_z$ as the displacements in radial
direction and parallel to the $z$-axis, respectively.

We consider the case that the molecule only exerts a total force
$\mathcal{F}$ perpendicular to the substrate surface:
\begin{eqnarray}
\mathcal{F}=k_M\left[x-\int_0^\infty\!\!\!  2\pi r f(r) u_z^0(r)dr\right], \label{totalforce}
\end{eqnarray}
where $u_z^0(r)$ is the parallel displacement at the surface
defined by $z=0$, and $f(r)$ is a normalized distribution
function, {\it i.e.} $\int 2\pi rf(r)dr =1$. This imposes the following
boundary conditions on the stress tensor
$T$:~\cite{footnoteVibPaper}
\begin{eqnarray}
\left.T_{zr} \right|_{z=0}=0,\qquad
\left.T_{zz} \right|_{z=0}=-\mathcal{F}f(r),
\end{eqnarray}
where the components $T_{zr}$ and $T_{zz}$ can be written as
functions of the displacements $u_r$ and $u_z$ (see, {\it e.g.},
Refs.~\onlinecite{landau59},~\onlinecite{kn:Ewing}). The solution
is then a straightforward generalization of the procedure for a
point source with $f(r)\propto\delta(r)/r$ outlined by Lamb in
Ref.~\onlinecite{lamb1904}. We obtain in frequency space
\begin{eqnarray}
u_z^0(r)&\!\!\!=&\!\!\! \mathcal{F}\frac{\omega^2}{v_t^2}
\int_0^\infty\frac{k\nu_lf_k}{ G(k,\omega)} J_0(kr)dk, \label{uz0}
\end{eqnarray}
where we have defined the following quantities:
\begin{eqnarray}
\nu_{t,l}&=&
\left\{\begin{array}{cl}
\sqrt{k^2-\frac{\omega^2}{v_{t,l}^2}}  & \mathrm{if}\,k^2\geq\frac{\omega^2}{v_{t,l}^2},\\
-i\sqrt{k^2-\frac{\omega^2}{v_{t,l}^2}} & \mathrm{if}\,k^2<\frac{\omega^2}{v_{t,l}^2},
       \end{array}
\right. \label{nutl}\\
G(k,\omega)&=&4\mu_L\nu_l\nu_t k^{2}-(\lambda_L+2\mu_L)(\nu_t^2+k^2)\nu_l^2\nonumber\\
&&\qquad +\lambda_L{k}^{2}\nu_t^{2}+\lambda_L{k}^{4}\label{denom},
\end{eqnarray}
where $\mu_L$ and $\lambda_L$ are the Lam\'{e} coefficients, which
are related to the sound velocities as
\begin{eqnarray}
v_l=\sqrt{\frac{\lambda_L+2\mu_L}{\rho}},\qquad v_t=\sqrt{\frac{\mu_L}{\rho}},
\end{eqnarray}
and $f_k$ is the Fourier-Bessel transform of the force distribution $f(r)$,
\begin{eqnarray}
f_k=\int_0^\infty f(r)J_0(kr) r dr. \label{fk}
\end{eqnarray}
The $(-)$-sign in the definition of $\nu_{t,l}$ in Eq. (\ref{nutl}) is necessary for
selecting the retarded response $\omega\rightarrow\omega+i\eta$
corresponding to outgoing waves since the square-root function has
a branch cut on the negative real axis.

The total force $\mathcal{F}$ involves $u_z^0(r)$ and vice versa,
see Eqs. (\ref{totalforce}, \ref{uz0}), so that we obtain:
\begin{eqnarray}
\int_0^\infty\!\!\!  2\pi r f(r) u_z^0(r)dr=x\frac{R(\omega)}{1+R(\omega)},
\end{eqnarray}
where
\begin{eqnarray}
R(\omega)=k_M \frac{2\pi \omega^2}{v_t^2}
\int_0^\infty\frac{k\nu_lf_k^2}{ G(k,\omega)} dk. \label{romega}
\end{eqnarray}
The function of interest, $\mysig(\omega)$, can then be deduced from the
equation of motion for the molecule in frequency space:
\begin{eqnarray}
-M\omega^2 x_0(\omega)=-k_M\left[x_0(\omega)-\int_0^\infty\!\!\!  2\pi r f(r) u_z^0(r)dr\right].
\end{eqnarray}
Identifying $k_M/\omega_M$ with the bare frequency $\omega_0^2$, we obtain
\begin{eqnarray}
\left[\omega^2-\omega_0^2+\omega_0^2\frac{R(\omega)}{1+R(\omega)}\right]x_0(\omega)=0,
\end{eqnarray}
which implies upon comparison with Eq.~(\ref{xeqmfreq})
\begin{eqnarray}
\mysig(\omega)=-\omega_0^2\frac{R(\omega)}{1+R(\omega)}\label{Sigma3D}.
\end{eqnarray}
For the situation where the molecule is attached to two substrates
it is simple to see that the function $\mysig(\omega)$ becomes
instead
\begin{eqnarray}
\mysig(\omega)=-\frac{k_{M,1}}{M}\frac{R_1(\omega)}{1+R_1(\omega)}
-\frac{k_{M,2}}{M}\frac{R_2(\omega)}{1+R_2(\omega)}\label{Sigma3D2},
\end{eqnarray}
where $R_{1,2}$ is given by Eq.~(\ref{romega}) but with $k_{M}$
replaced by $k_{M1,2}$ and, if the two substrates are different, with substrate
parameters changed accordingly. However, because of the lack of
detailed knowledge about the actual geometry of the device, and since the
coupling to the two sides of the junction is very likely to be
asymmetric, we will in the following make the simplifying
assumption that the molecule only couples to one substrate.
\begin{figure}[ptb]
\setlength{\unitlength}{1cm}
\begin{picture}
(6,5.5)(0,0) \put(-1,0){\includegraphics[width=8cm]{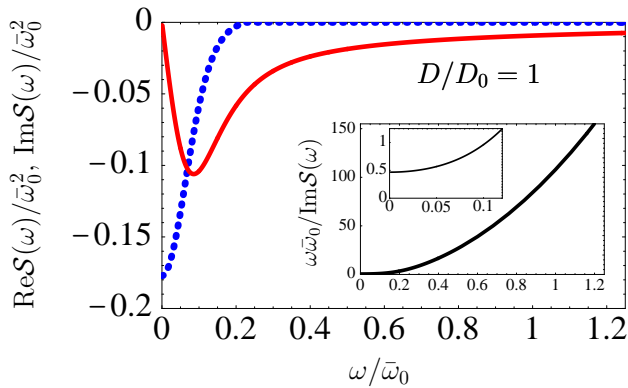}}
\end{picture}
\caption{Real part (dotted line) and imaginary part (solid line)
of $\mysig(\omega)$ as a function of frequency for a $C_{60}$ molecule on a gold
substrate and the particular choice  $D/D_0=1$. The imaginary part
tends towards zero linearly which is illustrated in the inset: the
``quality factor'' $\omega
\bar{\omega}_0/\mathrm{Im}\mysig(\omega)$ tends towards a constant
at zero frequency.} \label{fig:sigmaplots}
\end{figure}

Our results for $R(\omega)$ in Eq. (\ref{romega}) imply that the
imaginary part of $\mysig(\omega)$ (which will eventually be
responsible for the frictional damping) has contributions not only
from extended waves in the substrate but also from waves that are
confined to the surface, the so-called \emph{Rayleigh}
waves. Mathematically, this contribution arises
from $G(k,\omega)$ being zero for a specific value of $k$. This
value falls into the regime where both $\nu_{t}$ and $\nu_l$ are
real, {\it i.e.}, where wavevectors $k$ are larger than allowed
for transversal and longitudinal waves, see Eq. (\ref{nutl}).~\cite{lamb1904}

In order to compare our result (\ref{Sigma3D}) to experimental
data, we need to choose a specific model for the force
distribution function $f(r)$. The most realistic model would
involve a distribution in accord with the van-der-Waals potential,
however, as a result of that $f_k$ is a rather involved function
of $k$. For simplicity, we therefore choose
\begin{eqnarray}
f(r)=\frac1{2\pi D^2}\expe^{-r/D}\,, \,\mathit{i.e.},\,\,
f_k=\frac1{2\pi \sqrt{1+k^2D^2}^3}.\label{fkexpl}
\end{eqnarray}
The parameter $D$ is on the order of the width $D_0$ of the
molecule, {\it e.g.}, $D_0=$ 10.4\AA\,  for a $C_{60}$ molecule. For this
model, we can explicitly extract $\mysig(0)$, since then
\begin{eqnarray}
R(0)=\frac{3}{64}\frac{\omega_0^2 M \alpha^4}
{\left(\alpha^2-1\right)\rho v_l^2 D},\label{Romega0}
\end{eqnarray}
where $\alpha\equiv {v_l}/{v_t}$. Note that $R(0)$ is proportional
to the squared bare frequency
$\omega_0^2=\bar{\omega}_0^2-\mysig(0)$ so that we end up with
\begin{eqnarray}
\mysig(0)=-\bar{\omega}_0^2\frac{\bar{\omega}_0^2 R(0)
/\omega_0^2}{1+\bar{\omega}_0^2 R(0)/\omega_0^2}.
\end{eqnarray}
This result has the particular effect on the damping coefficient
$\mathrm{Im}\mysig(\omega)/\omega$ that it is independent of
$D/D_0$ at zero frequency. We show plots of the real and imaginary
parts of $\mysig(\omega)$ in Fig.~\ref{fig:sigmaplots}. The real
part, and thus the renormalization of the bare frequency as a
function of energy, goes to zero rather quickly, whereas the
imaginary part remains nonzero over a large frequency range. The
latter is important for the damping since the quantity $\omega
\bar{\omega}_0/\mathrm{Im}\mysig(\omega)$ takes the place of the
quality factor. The fact that this quantity tends towards a
constant at zero frequency illustrates that the imaginary part of
$\mysig(\omega)$ rises linearly with $\omega$ for small $\omega$.
\begin{figure}[ptb]
\setlength{\unitlength}{1cm}
\begin{picture}
(6,8.7)(0,0) \put(-1,0){\includegraphics[width=8cm]{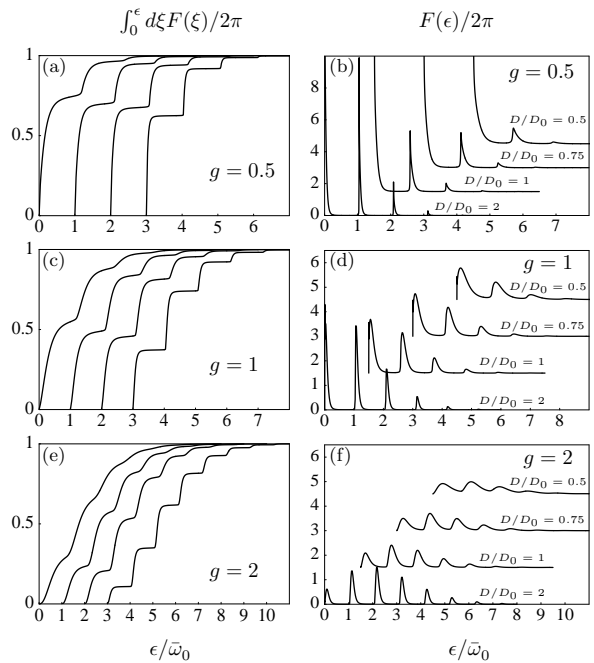}}
\end{picture}
\caption{The function $F(\xi)$ and its integral for different
values of $g$, making use of (\ref{Sigma3D}) for $\mysig(\omega)$. The curves have been
calculated from Eq.~(\ref{F1int}) at zero temperature, assuming a
$C_{60}$ molecule attached to a gold substrate. We take $D/D_0
=0.5$, 0.75, 1, 2, where $g=0.5$ in (a) and (b), $g=1$ in (c) and
(d), and $g=2$ in (e) and (f). The curves have been displaced for
clarity by multiples of (1,0) in (a), (c), and (e) (largest
$D/D_0$ to the right), and by multiples of (1.5,1.5) in (b), (d),
and (f) (largest $D/D_0$ at the bottom). The staircases in (a),
(c), and (e) feature sharper and less asymmetric steps for larger
$D/D_0$ but are clearly visible in any case. The asymmetry is even
more apparent in the plots of $F$ in (b), (d), and (f). In
contrast to the constant $Q$-factor results in
Fig.~\ref{fig:Qg052}, even the first step in the staircase gets
smeared for smaller $D/D_0$. However, we recover the large
constant-$Q$ limit for large $D/D_0$.} \label{fig:mysigFFInt}
\end{figure}

We plot the results for $F$ and its integral in
Fig.~\ref{fig:mysigFFInt}. First we note that the shape of the
staircases is markedly different from the constant $Q$-factor
model: they are asymmetric and less steep on the rising side with
a rather sharp transition to the next step. The asymmetry is even
more obvious in the peaks of $F$ itself. We also note that a
larger spread of the coupling over the surface, {\it i.e.} larger
$D/D_0$, makes the peaks in $F$ and the steps in its integral
sharper and less asymmetric. For large $D/D_0$, the staircase
tends towards the large constant $Q$ limit of before since then
$\omega/\mathrm{Im}\mysig(\omega)$ grows rapidly with $\omega$ and
at the same time $\mysig(0)\rightarrow 0$.

\section{$I$-$V$ curves}
\label{sec:iv}

\begin{figure}[ptb]
\setlength{\unitlength}{1cm}
\begin{picture}
(6,8)(0,0) \put(-1,0){\includegraphics[width=8cm]{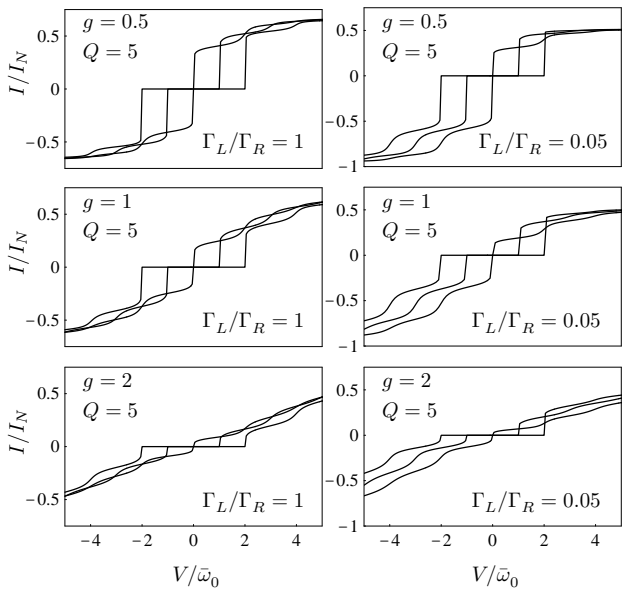}}
\end{picture}
\caption{Current-voltage characteristics for $g=0.5$, 1, 2, and
frequency-independent $Q=5$ at zero temperature. In each panel, we
have taken $\varepsilon_0=0$, $0.5\omega_0^{{}}$, and
$\omega_0^{{}}$, and the voltage is applied symmetrically across
the device, so that $V_{L}=V/2=-V_{R}$. The current is measured in
units of
$I_N=e\Gamma_L^{{}}\Gamma_R^{{}}(\Gamma_L^{{}}+\Gamma_R^{{}})$.
The panels on the left show the case of symmetric tunneling
contacts, whereas the panels on the right side correspond to
asymmetric tunneling contacts with $\Gamma_L/\Gamma_R=0.05$.}
\label{fig:iv1}
\end{figure}
\begin{figure}[ptb]
\setlength{\unitlength}{1cm}
\begin{picture}
(6,8)(0,0) \put(-1,0){\includegraphics[width=8cm]{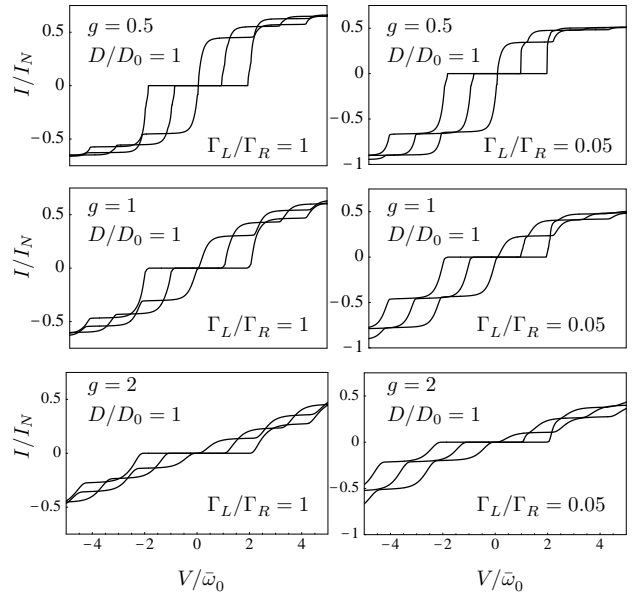}}
\end{picture}
\caption{Current-voltage characteristics for $g=0.5$, 1, and 2
using Eq. (\ref{Sigma3D}) for $\mysig(\omega)$ at zero
temperature, calculated for a $C_{60}$ molecule on gold. In each
panel, we have taken $\varepsilon_0=0$, $0.5\omega_0^{{}}$, and
$\omega_0^{{}}$, and the voltage is applied symmetrically across
the device, so that $V_{L}=V/2=-V_{R}$. The current is measured in
units of
$I_N=e\Gamma_L^{{}}\Gamma_R^{{}}(\Gamma_L^{{}}+\Gamma_R^{{}})$.
The panels on the left show the case of symmetric tunneling
contacts, whereas the panels on the right side correspond to
asymmetric tunneling contacts with $\Gamma_L/\Gamma_R=0.05$.}
\label{fig:ivmysigma}
\end{figure}

In this Section, we show a number of $I$-$V$ curves using the
expression in Eq.~(\ref{I}) at zero temperature based on the
$F$-function, both for the case of frequency-independent quality
factor and for the substrate model (\ref{Sigma3D}) for $\mysig(\omega)$ discussed
in the previous section.

In Fig.~\ref{fig:iv1}, we show current-voltage characteristics for
constant $Q=5$ and $g=0.5$, 1, 2. For this value of $Q$, the
Frank-Condon steps are still visible. If we took even smaller
values of $Q$ (not shown) such that the steps disappear, the
characteristics are still strongly modified by the electron-vibron
coupling in the sense that a gap develops in the $I$-$V$ curve.
Such an effect was recently claimed to be observed in a different
type of device.~\cite{hohb03}

We also show $I$-$V$ curves corresponding to a $C_{60}$ molecule
coupled to a gold substrate, using the substrate model of
Sec.~\ref{sec:realistic}, see Fig.~\ref{fig:ivmysigma}. We display
the $I$-$V$ curves for $g=0.5$, 1, 2, both for symmetric and for
asymmetric tunneling contacts, however, we restrict ourselves to
$D/D_0$=1 since the general features are very similar for other
choices of $D/D_0$. Upon comparison with the frequency-independent
$Q$-factor model, we note that the $I$-$V$ staircases are in
general less steep and smoother but still clearly exhibit the
expected Frank-Condon steps.

\section{Summary and discussion}
\label{sec:sumdis}

\subsection{Summary}
We have included broadening of the phonon sidebands due to
frictional coupling of the oscillator mode within a kinetic
equation approach. Since we have worked in the limit where the
tunneling time is much smaller than the lifetime of the
oscillator, we have assumed that the oscillator and the
environment are in thermal equilibrium and in this case an
analytical result for the current is obtained.

In the reference model featuring the frequency independent oscillator
quality factor $Q$, we recover the usual Frank-Condon physics for
large values of $Q$. The transition between the two different
charge states is then given by the usual overlap of two displaced
oscillator wavefunctions, the governing parameter being the ratio
of the displacement length $\ell$ and the oscillator length
$\ell_0^{{}}$, or $g=\ell^2/2\ell_0^2$. For moderate quality
factors $Q>Q_c^{{}}=2g/\pi$, the steps are smeared but still
visible. For even smaller values of the quality factor, the decay
time of the oscillations becomes shorter than the quantum
mechanical uncertainty time, which happens when $Q<1/g$. In this
strongly damped case the tunneling process crosses over to a
regime with a gap given by the classical displacement energy.

We were also able to calculate $\mysig(\omega)$ for a molecule
that is attached to a substrate and showed how the molecule loses
energy to the substrate. The model features similar general
results as the constant $Q$-factor model, however, it is different
in that the steps in the $I$-$V$ curves rise more smoothly but
feature a rather sharp transition to the next step, which then
again rises up smoothly. The underlying reason is the peak
structure of $F$, which exhibits asymmetric peaks due to the
frequency dependent damping coefficient. We also note the
dependence on the spread of the coupling over the substrate
surface, parameterized by $D/D_0$, where our results tend towards
the large constant $Q$ limit for large $D/D_0$.

\subsection{Comparison with experiments}
\begin{figure}[ptb]
\setlength{\unitlength}{1cm}
\begin{picture}
(6,4.7)(0,0) \put(-.5,0){\includegraphics[width=7cm]{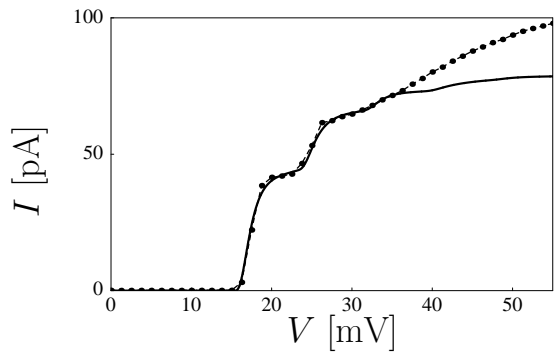}}
\end{picture}
\caption{ Example of a fit to the experimental curves of
Ref.~\onlinecite{park00} using the  substrate model (\ref{Sigma3D})
 for a $C_{60}$ molecule on gold, with $g=2$ and $D/D_0=0.75$.
The dots are experimental data points for a gate voltage of 6.8V
and positive bias voltage, and the solid line is the theoretical
curve. The smearing of the first step is seen to be reproduced
well, while at the same time showing a sharp rise for the second
step. This kind of smearing could not be produced by thermal
smearing, which would smear both steps equally. However, it is not
possible to make consistent fits for the entire $I$-$V$ curve and
for different gate voltages. This suggests that the molecule might
be changing position and/or coupling with changing voltages.}
\label{fig:exp}
\end{figure}

We have also tried to fit the present theoretical results to the
experiments in Ref.~\onlinecite{park00}, for which the theory
should be appropriate since the tunneling broadening is much
smaller than temperature, oscillator quantum, and observed widths. For
these experiments, in which $C_{60}$ molecules were attached to two
leads, it therefore seems likely that the broadening is dominated
by coupling to the environment.

A rough qualitative agreement, except for the steepness on the
rising side of the steps, can be achieved for the
frequency-independent quality factor if we assume $g\approx 1-2$
and $Q\approx 2-6$. However, in order to obtain quantitative
agreement, it is necessary to assume different values for $g$ and
$Q$ for different values of the gate and source-drain voltages.

Our model for $\mysig(\omega)$ that corresponds to a molecule
attached to a substrate, see Sec.~\ref{sec:realistic}, features
qualitative agreement with experiment if we assume $g$ and $D/D_0$
to be on the order of unity. The asymmetry in the peak structure
of $F$ actually provides for a better quantitative fit to the
experimental data than is possible for the constant $Q$-factor
model. This is illustrated in Fig.~\ref{fig:exp}.

\acknowledgments The authors acknowledge discussions with
P.~Brouwer, A.-P. Jauho, P.~McEuen, T.~Novotny, J.~Park, and
J.~Sethna. The work was supported by the Cornell Center
for Materials Research under NSF Grant No.~DMR0079992, by the Danish National Research
Council, and by the Packard foundation.


\begin{thebibliography}{10}

\bibitem{reed97}
M.~A. Reed {\it et~al.}, Science {\bf 278},  252  (1997).

\bibitem{park00}
H. Park {\it et~al.}, Nature {\bf 407},  57  (2000).

\bibitem{park02}
J. Park {\it et~al.}, Nature {\bf 417},  722  (2002).

\bibitem{lian02}
W. Liang {\it et~al.}, Nature {\bf 417},  725  (2002).

\bibitem{smit02}
R.~H.~M. Smit {\it et~al.}, Nature {\bf 419},  906  (2002).

\bibitem{zhit02}
N.~B. Zhitenev, H. Meng, and Z. Bao, Phys. Rev. Lett. {\bf 88},
226801
  (2002).

\bibitem{boes01}
D. Boese and H. Schoeller, Europhys. Lett. {\bf 54},  668  (2001).

\bibitem{mcca02}
K. McCarthy, N. Prokof'ev, and M. Tuominen, cond-mat/0205419.

\bibitem{mitr03}
A. Mitra, I. Aleiner, and A.~J. Millis, cond-mat/0302132.

\bibitem{aji03}
V. Aji, J. Moore, and C. Varma, cond-mat/0302222.

\bibitem{gure98}
V.~L. Gurevich and H.~R. Schober, Phys. Rev. B {\bf 57},  11295
(1998).

\bibitem{patt02}
K.~R. Patton and M.~R. Geller, cond-mat/0202325.

\bibitem{naza89a}
Y.~V. Nazarov, Zh. Eksp. Teor. Fiz. {\bf 95},  975  (1989), [Sov.
Phys. JETP
  {\bf 68}, 561 (1989)].

\bibitem{devo90}
M.~H. Devoret {\it et~al.}, Phys. Rev. Lett. {\bf 64},  1824
(1990).

\bibitem{girv90}
S.~M. Girvin {\it et~al.}, Phys. Rev. Lett. {\bf 64},  3183
(1990).

\bibitem{ingo91}
G.-L. Ingold and H. Grabert, Europhys. Lett. {\bf 14},  371
(1991).

\bibitem{cald83}
A.~O. Caldeira and A.~J. Leggett, Ann. Phys. {\bf 149},  374
(1983).

\bibitem{ard95}
A.~A. Louis and J.~P. Sethna, Phys. Rev. Lett. {\bf 74},  1363
(1995).

\bibitem{mahan}
G.~D. Mahan, {\em Many-Particle Physics} (Plenum Press, New York,
1990).

\bibitem{alex02}
A. Alexandrov and A. Bratkovsky, Phys. Rev. B {\bf 67},  235312
(2003).

\bibitem{minn76}
P. Minnhagen, Phys. Lett. {\bf 56A},  327  (1976).

\bibitem{kn:ezawa}
H. Ezawa, Ann. Phys. {\bf 67},  438  (1971).

\bibitem{footnoteVibPaper}
The stress tensor is defined as $d\vec{F}=Td\vec{A}$, where
$d\vec{F}$ is an
  infinitesimal force, and $d\vec{A}$ is an infinitesimal area element.

\bibitem{landau59}
L.~D. Landau and E.~M. Lifshitz, {\em Theory of Elasticity}
(Pergamon Press,
  New York, 1959).

\bibitem{kn:Ewing}
W.~M. Ewing, W.~S. Jardetzky, and F. Press, {\em Elastic Waves in
Layered
  Media} (McGraw-Hill, New York, 1957).

\bibitem{lamb1904}
H. Lamb, Phil. Trans. Roy. Soc. A {\bf 203},  1  (1904).

\bibitem{hohb03}
E.~M. H\"ohberger {\it et~al.}, cond-mat/0304136.

\end{thebibliography}
\end{document}